\documentclass[12pt]{article}
\usepackage{graphicx}
\usepackage{amssymb}
\usepackage{amsmath, amsthm}
\newcommand{\be}{ \begin{equation} }
\newcommand{\ee}{\end{equation}} 
\begin{document}  
\def\theequation{\arabic{section}.\arabic{equation}}
\begin{titlepage}
\title{Horizons and singularity in Clifton's spherical solution
of $f(R)$ vacuum}
\author{Valerio Faraoni \\ \\
{\small \it Physics Department, Bishop's University}\\
{\small \it 2600 College St., Sherbrooke, Qu\'{e}bec, Canada 
J1M~1Z7}
}
\date{} \maketitle
\thispagestyle{empty}
\vspace*{1truecm}
\begin{abstract}
Due to the failure of Birkhoff's theorem, black holes in 
$f(R)$ gravity theories in which an effective time-varying
cosmological ``constant'' is present are, in general, 
dynamical. Clifton's exact spherical solution of $R^{1+\delta}$ 
gravity, which is dynamical and describes a central 
object embedded in  a  spatially flat universe, is studied. It is 
shown that apparent black hole horizons disappear and a naked  
singularity emerges at late times.
\end{abstract}
\end{titlepage} 
\clearpage 
\setcounter{page}{2} 

\section{Introduction}

The study of Type Ia supernovae \cite{SN1,SN2, SN3, SN4, SN5, 
SN6, SN7, SN8, SN9} revealed that  the 
universe is currently accelerating its expansion. 
This discovery has generated an enormous amount of activity and 
theoretical  models in order to find an explanation of this 
phenomenon. The most common models are based on  
General  Relativity (GR) and invoke  mysterious 
forms of  dark energy  (see  \cite{Linderresletter} for a list of 
references). However, dark energy, possibly even  phantom energy,  
is too exotic and {\em ad hoc} and attempts have been made to 
model the cosmic  acceleration without dark energy. $f(R)$ 
theories of gravity reminiscent of the quadratic corrections to 
the  Einstein-Hilbert action introduced by renormalization have 
been re-introduced in the metric  \cite{CCT, CDTT}, Palatini 
\cite{Vollick}, 
and metric-affine \cite{metricaffine1, metricaffine2, 
metricaffine3, metricaffine4, metricaffine5} formulations and 
have 
received  much attention in recent years (see \cite{review, 
DeFeliceTsujikawa10} for  reviews  and \cite{otherreviews1, 
otherreviews2, otherreviews3, otherreviews4, otherreviews5, 
otherreviews6, otherreviews7} for  introductions). Emilio 
Elizalde has given many contributions to the development of 
$f(R)$ gravity and cosmology.

While many cosmological and other aspects of $f(R)$ gravity  
(stability, weak-field limit, ghost content) have  been  
discussed in recent years, it is important to understand 
spherical solutions (both vacuum and interior) in 
these theories \cite{Frolovetc1, Frolovetc2, Frolovetc3, 
Frolovetc4, Frolovetc5, Frolovetc6, Frolovetc7, Frolovetc8, 
Frolovetc9}.

Metric $f(R)$ gravity is described by the action
\begin{equation}
S=\frac{1}{2\kappa} \int d^4x \sqrt{-g}\, f(R) +S^{(matter)} \,,
\end{equation}
where $f(R)$ is a non-linear function of its argument and  
$S^{(matter)} $ is the matter part of the action. $R$ 
denotes  the Ricci scalar of the metric $g_{ab}$ with determinant 
$g$, $\kappa \equiv 8\pi G$ where $G$ is Newton's constant, and 
we adopt  the notations of Ref.~\cite{Wald}. 

The Jebsen-Birkhoff  theorem of GR fails in these 
theories, adding to the  variety of spherical solutions 
\cite{BirkhoffPRD}. Of 
particular interest are black holes in generalized gravity, which 
have been studied especially in relation to their  
thermodynamics (\cite{myEntropy} and references therein).  
Since $f(R)$ theories  are designed to produce an  effective 
dynamical cosmological constant, physically relevant spherically 
symmetric and black hole solutions are likely to describe
central objects embedded in 
cosmological 
backgrounds. This kind of solution is poorly understood  
even in GR, although a few examples are available there  
\cite{SultanaDyer1, SultanaDyer2, SultanaDyer3, 
SultanaDyer4, McClureDyer1, McClureDyer2, FaraoniJacques,  
SaidaHaradaMaeda, Gaoetal, Hideki, 
HidekiHaradaCarr, fate}. Even less is known 
about $f(R)$ black holes, which are certainly worth exploring. 
Here we consider a specific solution of vacuum 
$f(R)=R^{1+\delta}$ gravity discovered in \cite{Clifton}.
Solar System experiments  set the  limits\footnote{See also 
\cite{CordaMosquera} for this specific form of the function 
$f(R)$.}  
$\delta 
=\left( -1.1\pm 1.2 \right) \cdot 10^{-5}$ on the parameter $ 
\delta$ \cite{Clifton, BarrowClifton1, BarrowClifton2, 
BarrowClifton3, BarrowClifton4}, while local 
stability requires $f''(R)  \geq 0$ \cite{DolgovKawasaki, 
mattmodgrav1, mattmodgrav2, Odintsovconfirm},  hence we restrict 
to the range 
$0<\delta < 
10^{-5}$. 

The solution of \cite{Clifton} is dynamical and describes 
a time-varying central object embedded in a 
spatially flat universe in vacuum $ R^{1+\delta}$ 
gravity.
This solution is made possible by the fact that  
the fourth order field equations of  metric $f(R)$ gravity
\begin{equation}
f'(R)R_{ab}-\frac{f(R)}{2}\, g_{ab}=\nabla_a\nabla_b 
f'(R)-g_{ab} \Box f'(R) 
\end{equation}
in vacuo can be rewritten as effective Einstein 
equations 
\begin{equation} 
R_{ab}-\frac{1}{2}\, g_{ab} R=\frac{1}{f'(R)}\left[ 
\nabla_a\nabla_b f'-g_{ab} \Box f' +g_{ab}\frac{\left( 
f-Rf'\right)}{2} \right] 
\end{equation}
with  geometric terms acting as effective  matter on 
the right hand side. This time-varying effective matter 
invalidates the Jebsen-Birkhoff 
theorem and can propel the acceleration of the universe. 
An equivalent representation of metric $f(R)$ gravity as an 
$\omega=0$  Brans-Dicke  theory with a special scalar field 
potential reveals explicitly  the presence of a massive scalar   
degree of freedom $f'(R)$ responsible for these effects 
\cite{review}. 
Since analytical spherical and dynamical solutions of $f(R)$ 
gravity in asymptotically Friedmann-Lemaitre-Robertson-Walker 
(FLRW) backgrounds are harder to find 
than in  GR (where only a few are known anyway), Clifton's 
solution is particularly valuable.

\section{Clifton's solution and its horizons}
\setcounter{equation}{0}

In this section we describe the Clifton solution \cite{Clifton} 
and the work \cite{myClifton} locating the horizons of this 
solution.

The spherically symmetric and time-dependent  solution of vacuum 
$ R^{1+\delta}$ gravity of \cite{Clifton} is given by  
\begin{equation}\label{1}
ds^2=-A_2(r)dt^2+a^2(t)B_2(r)\left[ dr^2 +r^2 \left(  d\theta^2 
+\sin^2 \theta \, d\varphi^2  \right) \right] 
\,,
\end{equation}
and (using the isotropic radius and the notations of 
\cite{Clifton})
\begin{eqnarray}
A_2(r) &=& \left( \frac{1-C_2/r}{1+C_2/r}\right)^{2/q} \,, 
\label{2} \\
&&\nonumber \\
B_2(r) &=& \left( 1+\frac{C_2}{r} \right)^{4}A_2(r)^{\, q+2\delta 
-1} \,,\label{4}\\
&&\nonumber \\
a(t) &= & t^{\frac{ \delta \left( 1+2\delta \right)}{1-\delta}} 
\,,\label{3}\\
&&\nonumber\\
q^2 &= & 1-2\delta+4\delta^2 \,. \label{5}
\end{eqnarray}
Once $\delta $ is fixed, two classes of solutions exist, 
corresponding to the sign of  $C_2 q r $.  The line 
element~(\ref{1}) reduces to the FLRW one if  $C_2\rightarrow 0$. 
In the  limit $\delta \rightarrow 0$ in which the theory reduces to 
GR, eq.~(\ref{1}) reduces to the 
Schwarzschild metric in isotropic coordinates provided that 
$C_2 q r>0$, hence positive and negative values of $r$ are 
possible according to the sign of $C_2$, but we assume  
$r>0, C_2>0$ and take the positive root in the expression $q=\pm 
\sqrt{1-2\delta +4\delta^2}$, so 
that $ q\simeq 1-\delta$ as $ \delta \rightarrow 0 $.  
The  solution~(\ref{1})-(\ref{5}) is conformal to the 
Fonarev  solution \cite{Fonarev}  which is conformally static 
\cite{HidekiFonarev}, and therefore is also conformally static, 
similar to the  Sultana-Dyer \cite{SultanaDyer1, SultanaDyer2, 
SultanaDyer3,  SultanaDyer4} and certain  
generalized McVittie  solutions \cite{fate} of GR.

In order to identify possible apparent horizons, it is 
convenient 
to cast the metric~(\ref{1}) in the Nolan gauge. Using first  
the  Schwarzschild-like radius 
\begin{equation} \label{6}
\tilde{r} \equiv r \left( 1+\frac{C_2}{r} \right)^2 \,,
\end{equation}
giving $dr=\left( 1-\frac{C_2^2}{r^2} \right)^{-1} 
d\tilde{r} $ and then  the areal radius
\begin{equation}\label{8}
\rho \equiv \frac{ a(t) \sqrt{B_2(r)} \, \tilde{r} }{\left( 
1+\frac{C_2}{r} \right)^2} =a(t) \, \tilde{r} \, A_2(r)^{ 
\frac{q+2\delta -1}{2}}  \,,
\end{equation}
the line element~(\ref{1}) takes the form
\begin{equation}\label{9}
ds^2=-A_2dt^2 +a^2 A_2^{2\delta -1}d\tilde{r}^2 +{\rho}^2 
d\Omega^2  \,.
\end{equation}
Denoting the differentiation with respect to time with an overdot 
and using the identities
\begin{equation} \label{10}
d\tilde{r}=\frac{
d\rho-A_2^{\frac{q+2\delta -1}{2} } \dot{a} \, \tilde{r} \, dt}{
a\left[ A_2^{\frac{q+2\delta -1}{2}} +\frac{2( q+2\delta -1)}{q} 
\frac{C_2}{\tilde{r}} A_2^{\frac{2\delta -1-q}{2}} \right]} 
\equiv 
\frac{ d\rho -A_2^{\frac{q+2\delta-1}{2}} \dot{a} \, \tilde{r} \, 
dt 
}{a 
A_2^{\frac{q+2\delta-1}{2}} C(r)} \,,
\end{equation}
one obtains 
\begin{equation}\label{11}
C(r)=1+\frac{2(q+2\delta -1)}{q} \, \frac{C_2}{\tilde{r}} \, 
A_2^{-q} = 
1+\frac{2(q+2\delta -1)}{q} \, \frac{C_2 a}{\rho} \, 
A_2^{\frac{2\delta -1-q}{2}} \,,
\end{equation}
which turns the metric into the Painlev\'e-Gullstrand-like 
form
\begin{eqnarray}
ds^2 & = & - A_2\left[ 1-\frac{A_2^{ 2(\delta -1)} }{C^2}\, 
\dot{a}^2 \tilde{r}^2 \right] 
dt^2-\frac{2A_2^{\frac{-q+2\delta-1}{2}}}{C^2} \, 
\dot{a} \, \tilde{r} \, dtd\rho \nonumber\\
&&\nonumber\\
&+ &  \frac{d\rho^2}{A_2^q C^2}+ \rho^2d\Omega^2 \,.\label{12}
\end{eqnarray}
Now we introduce a new time coordinate $\bar{t}$ defined by  
\begin{equation}\label{13}
d\bar{t}=\frac{1}{F(t, \rho)} \left[ dt +  \beta(t,\rho)d\rho 
\right] 
\end{equation}
in order to eliminate the cross-term $dtd\rho$. Here $F(t,\rho)$ 
is an integrating factor which guarantees that $d\bar{t}$ is an 
exact differential and is determined by 
\begin{equation}\label{14}
\frac{\partial}{\partial \rho}\left( \frac{1}{F} \right)=
\frac{\partial}{\partial t}\left( \frac{\beta}{F} \right) \,.
\end{equation}
The line element  becomes 
\begin{eqnarray}
ds^2 & = & -A_2\left[ 1-\frac{A_2^{ 2(\delta -1)} }{ 
C^2}\,  \dot{a}^2 \tilde{r}^2 \right] F^2 d\bar{t}^2 
\nonumber\\
&&\nonumber\\
 &+ & 2F  
\left\{ A_2\beta \left[  1-\frac{A_2^{ 2(\delta -1)} }{C^2}
\, \dot{a}^2 \tilde{r}^2 \right]
-\frac{A_2^{\frac{-q+2\delta-1}{2}} }{C^2} \, 
\dot{a}  \tilde{r} \right\}
d\bar{t}d\rho \nonumber\\
&&\nonumber\\
& + & \left\{-A_2 
\left[ 1-\frac{A_2^{ 2(\delta -1)} }{C^2}\, 
\dot{a}^2  \tilde{r}^2 \right]\beta^2 
+\frac{2A_2^{\frac{-q+2\delta -1}{2}}}{C^2}\, 
\dot{a}\tilde{r}\beta +\frac{1}{A_2^qC^2} \right\} 
d\rho^2 \nonumber\\
&&\nonumber\\ 
& + &\rho^2d\Omega^2 
\,.\label{15}
\end{eqnarray}
The choice
\begin{equation}\label{16}
\beta= \frac{ A_2^{ \frac{-q+2\delta-3}{2} } }{C^2}\, 
  \frac{\dot{a} \, \tilde{r}}{  1-
\frac{A_2^{2(\delta -1)} }{C^2}\, 
\dot{a}^2 \tilde{r}^2 }  
\end{equation}
eliminates the $dtd\rho$ term and casts the metric in the 
Nolan gauge 
\begin{eqnarray}
ds^2 & = & -A_2 D F^2 d\bar{t}^2 +\frac{1}{A_2^q C^2} \left[
1+\frac{ A_2^{-q-1} H^2 
\rho^2}{C^2 D} 
\right] d\rho^2 \nonumber\\
&&\nonumber\\
&+ & \rho^2 d\Omega^2 \,, \label{18}
\end{eqnarray}
where $ H\equiv \dot{a}/a$ is the Hubble parameter of the 
background universe and 
\begin{equation}\label{17}
D\equiv 1-\frac{ A_2^{2(\delta -1)} }{C^2}\, 
\dot{a}^2\tilde{r}^2 =
1-  \frac{A_2^{-q-1} }{C^2}\, 
H^2 \rho^2 \,.
\end{equation}
Using the second of these equations, the line 
element~(\ref{18}) assumes the simple form
\begin{equation}
ds^2 = -A_2 D F^2 d\bar{t}^2 +\frac{d\rho^2}{A_2^q C^2 D} 
+  \rho^2 d\Omega^2 \,.
\end{equation}
The apparent horizons, if they exist, are located at 
$g^{\rho\rho}=0$, which yields  $  A_2^q C^2 D =0$ 
and  $ A_2^q \left( C^2 -H^2R^2 A_2^{-q-1} \right)=0 $. 
Therefore,  $g^{\rho\rho}$ vanishes if $A_2=0$ or $H^2R^2=C^2 
A_2^{q+1}$.  $A_2$ vanishes at  $ r=  C_2 $, which 
describes the  Schwarzschild  horizon when 
$\delta\rightarrow  0$ (the GR limit). This locus corresponds  
to a  spacetime singularity because the Ricci 
scalar $  R=\frac{ 6\left( \dot{H}+2H^2 \right)}{A_2(r)} $ 
diverges as $r\rightarrow C_2$ (it reduces  to the usual FLRW 
value $6\left( \dot{H}+2H^2 \right)$ as  $C_2\rightarrow 0$). 
This singularity is strong according to Tipler's classification 
\cite{Tipler} because the areal radius  $ \rho =a\tilde{r} 
A_2^{\frac{q+2\delta-1}{2} } $ vanishes when 
$r=C_2 $  for $\delta >0$,  in contrast with the  Schwarzschild 
metric corresponding to $\delta=0$ in which $ \rho 
=\tilde{r}=4C_2$  at  $r=C_2$.

The second possibility $H^2 \rho^2=C^2 
A_2^{q+1}$ yields 
\begin{equation} \label{24}
H \rho =\pm\left[ 1+\frac{2(q+2\delta -1)}{q}\, \frac{C_2 
a}{\rho}\, 
A_2^{\frac{2\delta-1-q}{2}} \right] A_2^{\frac{q+1}{2}} \,,
\end{equation}
with the positive sign corresponding to an expanding 
universe. When $\delta \rightarrow 0$, this equation reduces 
to $ H  \rho =  
\left[ 
1+\frac{2\delta C_2 a}{ \rho}\, A_2^{-\left(1-\frac{3\delta}{2} 
\right)} \right] A_2^{1-\delta} $. 

To gain some insight, consider the following two limits. As  
$C_2\rightarrow  0$ (the central object disappears 
and the solution is FLRW space), $r=\tilde{r}$  and $ \rho$  
become a comoving and a proper radius, respectively, while 
eq.~(\ref{24}) reduces to $H\rho=1$ with solution $ 
\rho_{c}=1/H$, the radius of the cosmological horizon.  In the 
limit $\delta\rightarrow 0 $ in  which the theory reduces to GR, 
eq.~(\ref{24}) reduces to $A_2=0$  or $r=C_2$ with $H\equiv 0$.

Using eqs.~(\ref{3}) and (\ref{8}), the left hand side of 
eq.~(\ref{24}) is expressed as
\begin{equation}
HR=\frac{ \delta \left( 1+2\delta \right)}{1-\delta}\, t^{ 
\frac{2\delta^2+2\delta -1}{1-\delta}} \, \frac{C_2}{x} \, \frac{ 
\left( 1-x \right)^{ \frac{ q+2\delta-1}{q}}}{\left( 1+x\right)^{ 
\frac{-q+2\delta-1}{q}}} \,,
\end{equation}
where $x \equiv C_2/r$, while the right hand side of~(\ref{24})  is
\begin{equation}
\left( \frac{1-x}{1+x} \right)^{\frac{q+1}{q} }  \left[ 
1+\frac{2\left( 
q+2\delta-1\right)}{q} \, \frac{x}{\left( 1-x\right)^2} \right] 
\,.
\end{equation}
Eq.~(\ref{24}) then becomes
\begin{eqnarray}
 \frac{1}{t^{ \frac{1-2\delta-2\delta^2}{1-\delta} } } 
& = & \frac{\left( 1-\delta\right)}{\delta\left( 
1+2\delta\right)C_2}\, 
\frac{ x \left( 1+x\right)^{ \frac{ -2q+2\delta-2}{q} }}{ \left( 
1-x\right)^{\frac{2\left( \delta-1\right)}{q}}} \nonumber\\
&&\nonumber\\
& \cdot & 
\left[ 1+\frac{2\left( q+2\delta -1 \right)}{q}\, 
\frac{x}{(1-x)^2} \right] \label{DELTA}
\end{eqnarray}
(note that $\frac{1-2\delta- 2\delta^2}{1-\delta}$ is positive 
for $0<\delta < \frac{\sqrt{3} \, -1}{2} \simeq 0.366$).

\begin{figure}[b]
%\sidecaption
% For example, with the graphicx style use
\includegraphics[width=9.5cm]{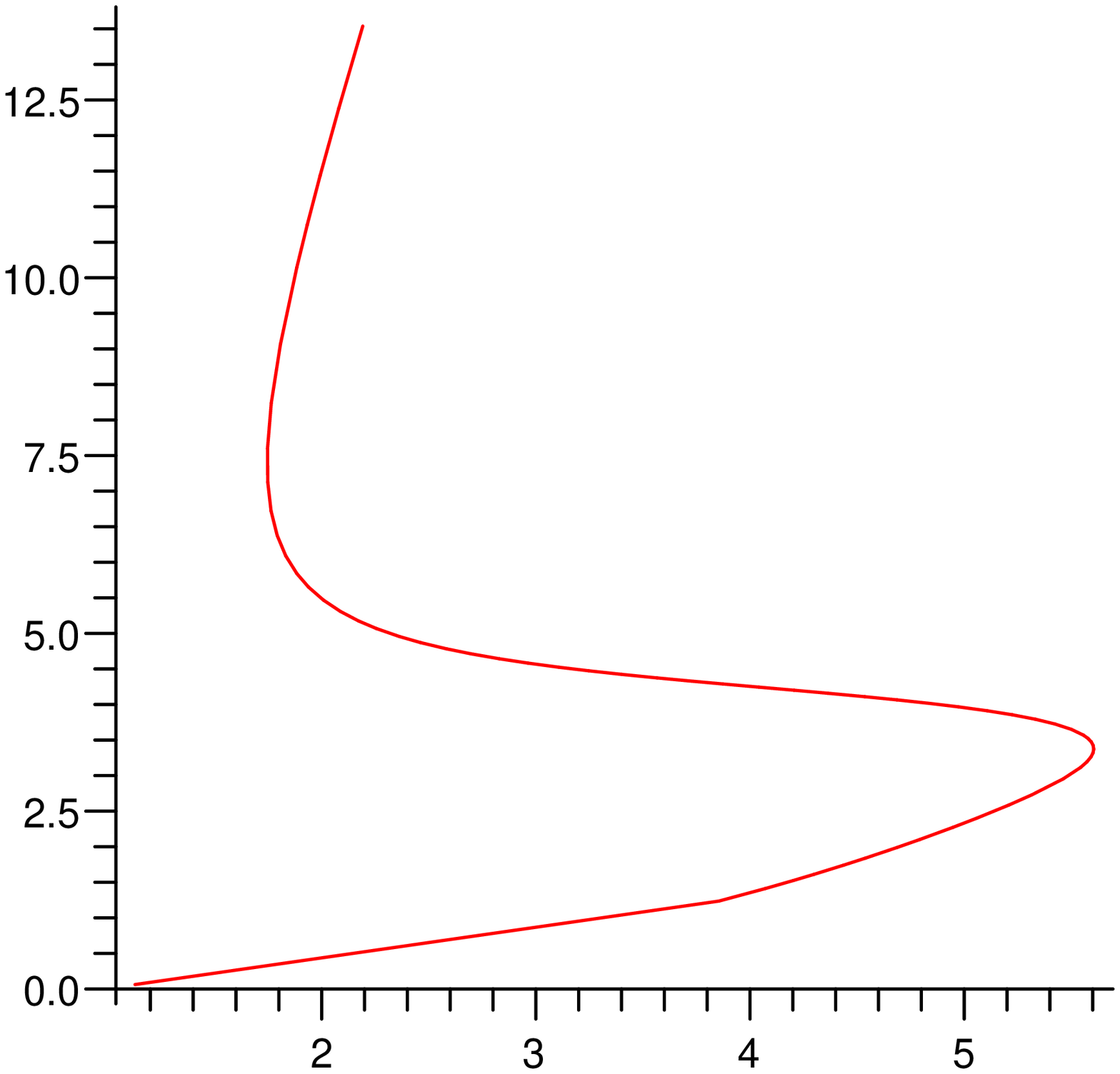}
%\caption{If the width of the figure is less than 7.8 cm  use the 
%\texttt{sidecapion} command to flush the caption on the  left 
%side of the page. If the figure is positioned at the top  of the 
%page, align the sidecaption with the top of the figure --  to 
%achieve this you simply need to use the optional argument  
%\texttt{[t]} with the \texttt{sidecaption} command}
\label{fig:1}
\caption{Radii of the apparent horizons of Clifton's solution 
(vertical axis) versus time (horizontal axis) for the  
parameter values $C_2=1$ and $\delta=0.13$.}       
\end{figure}

At late times $t$, the left hand side of eq.~(\ref{DELTA}) 
vanishes, $x\simeq 0$, and  there exists a unique root of the 
equation locating the apparent  horizons, which corresponds to  
a cosmological horizon, consistently with the fact that 
$r\rightarrow \infty$ as  $x=C_2/r \rightarrow 0$. The limit 
$x\rightarrow 0$ can also be obtained when  the  
parameter $C_2 \rightarrow 0$ , in which case $H\rho \rightarrow 
1$ and  $r\simeq \rho \simeq H^{-1}=\frac{1-\delta}{\delta \left( 
1+2\delta \right)}\, t$ is the radius of the cosmological horizon 
of the FLRW space without a  central object. Hence,  {\em there 
is only a cosmological apparent horizon and no black hole 
apparent horizons at late times}: the central 
singularity  at $\rho=0$ becomes naked.

The radii $\rho$ of the apparent horizons and the time $t$  
can  be expressed in the parametric form 
\begin{eqnarray}
\rho(x) &=& t(x)^{ \frac{\delta \left(1+2\delta 
\right)}{1-\delta}}  \frac{C_2}{x} \, \left( 
1-x\right)^{\frac{q+2\delta-1}{q}} \left( 
1+x \right)^{\frac{q-2\delta+1}{q}} \,,\\
&&\nonumber\\
t(x) &=&   \left\{ \frac{\left( 1-\delta\right)}{\delta\left( 
1+2\delta\right)C_2}\, 
\frac{ x \left( 1+x \right)^{ \frac{ 2\left( 
-q+\delta-1\right)}{q} }}{ 
\left(  1-x\right)^{\frac{2\left( \delta-1\right)}{q}}}
\left[ 1+\frac{2\left( q+2\delta -1 \right)x}{q(1-x)^2} \right] 
\right\}^{\frac{1-\delta}{2\delta^2+2\delta-1}} \,,
\label{questa} 
\end{eqnarray}
using  $x$ as  a parameter.  Fig.~1 reports $\rho$ versus $t$  
for  the parameter values $C_2=1$ and $\delta=0.13$, showing that  
two inner horizons develop  after the Big Bang covering the 
central singularity $\rho=0$, then they approach each other, 
merge, and  disappear, while a third,  
cosmological horizon keeps expanding.  The $\rho=0$ singularity 
becomes naked after this merging event.

\section{Discussion and conclusions}
\setcounter{equation}{0}

Cosmologists  may be detecting  deviations from GR 
and therefore it is necessary to understand  spherical 
solutions of $f(R)$ gravity, which has been proposed as a simple 
alternative to the mysterious dark energy.  Since the 
Jebsen-Birkhoff  theorem fails in these theories, spherical  
solutions do not have 
to be static. $f(R)$ theories are designed with a built-in 
dynamical cosmological constant to model the present acceleration 
of the universe, hence analytical spherical solutions 
describing a central object embedded in a FLRW background are the 
relevant ones.  Unfortunately, such solutions are poorly 
understood   even in GR  \cite{SultanaDyer1, SultanaDyer2, 
SultanaDyer3, 
SultanaDyer4, McClureDyer1, McClureDyer2, 
FaraoniJacques,  SaidaHaradaMaeda, Gaoetal, Hideki, 
HidekiHaradaCarr, fate}. It seems difficult to 
find {\em  generic} solutions describing black holes 
embedded in FLRW backgrounds. Finding  numerically spherical 
interior solutions of $f(R)$ gravity is also an active area of 
research  \cite{Frolovetc1, Frolovetc2, Frolovetc3, 
Frolovetc4, Frolovetc5, Frolovetc6, Frolovetc7, Frolovetc8, 
Frolovetc9}. All these issues deserve further 
attention in the future.

\section*{Acknowledgments}

It is a pleasure to thank  Rituparno Goswami for a discussion, 
the conference organizers for their invitation,  and the 
Natural Sciences and Engineering Research Council of Canada 
(NSERC) for financial support.

\vskip1truecm
%\clearpage

\end{document}